\documentclass[]{aa}
\usepackage{txfonts}
\usepackage[utf8x]{inputenc}
\usepackage{graphicx}
\usepackage{natbib}
\usepackage{epstopdf}
\bibpunct{(}{)}{;}{a}{}{,} 

\begin{document}

\def\simgt{\lower.5ex\hbox{$\; \buildrel > \over \sim \;$}}
\def\simlt{\lower.5ex\hbox{$\; \buildrel < \over \sim \;$}}
\newcommand\igr{IGRJ17480} 
\newcommand\VMI{V--I} 
\newcommand{\msec}{$M_{2}$} 
\newcommand{\mprim}{$M_{1}$} 
\newcommand\teff{$T_{\rm eff}$}
\newcommand\dm{$D_{\rm mix}$}
\newcommand\Teff{$T_{\rm eff}$}
\newcommand\Mv{$M_{\rm V}$}
\newcommand\dlt{\Delta\log\,T}
\newcommand\cs{\,cm$^2$\,s$^{-1}$}
\newcommand\kms{\,km\,s$^{-1}$}
\newcommand\ks{$K_\mathrm{S}$}
\newcommand{\msun}{\ensuremath{\, {M}_\odot}}
\newcommand{\Msun}{\ensuremath{\, {M}_\odot}}
\newcommand{\ocen}{$\omega$~Cen}
\newcommand{\mlate}{M$_{mixing}$}
\def\he3{$^3$He}
\def\vmi{\hbox{\it V--I\/}}                                 
\def\bmv{\hbox{\it B--V\/}}                                 
\def\bmi{\hbox{\it B--I\/}}                                 

\title{The near-IR counterpart of \object{IGR J17480-2446} in \object{Terzan 5}\thanks{Based on observations collected at the European Organisation for Astronomical Research in the Southern Hemisphere, Chile, DDT proposal 286.D-5012. 
Based on observations made with the NASA/ESA Hubble Space Telescope, obtained from the data archive at the Space Telescope Science Institute. 
STScI is operated by the Association of Universities for Research in Astronomy, Inc. under NASA contract NAS 5-26555.} }
\author{V. Testa\inst{1} \and T. di Salvo\inst{2} \and F. D'Antona\inst{1} \and M.T. Menna\inst{1} \and P. Ventura\inst{1} \and L.Burderi\inst{3} \and A. Riggio\inst{3} \and R. Iaria\inst{2} 
\and A. D'A\`\i\inst{2} \and A. Papitto\inst{4} \and N. Robba\inst{2}}

\institute{INAF-Osservatorio Astronomico di Roma, Via Frascati, 33 - 00040 Monte Porzio Catone - Italy \and
Dipartimento di Fisica, Universit\`a di Palermo, Via Archirafi 36, 90123 Palermo -Italy \and
Universit\`a di Cagliari, Dipartimento di Fisica, SP Monserrato-Sesto km 0.7, 09042 Monserrato (CA) - Italy \and
Institut de C{\`\i}encies de l'Espai (IEEC-CSIC), Campus UAB, Fac. de C{\`\i}encies, Torre C5, parell, 2a planta, 08193 Barcelona, Spain}


\abstract {Some globular clusters in our Galaxy are noticeably rich in low-mass X-ray binaries. \object{Terzan 5} has the richest population among globular clusters 
of X- and radio-pulsars and low-mass X-ray binaries.} 
{The detection and study of optical/IR counterparts of low-mass X-ray binaries is fundamental to characterizing both the low-mass donor in the binary 
system and investigating the mechanisms of the formation and evolution of this class of objects. We aim at identifying the near-IR counterpart of the 11 Hz 
pulsar \object{IGRJ17480-2446} discovered in \object{Terzan 5}.} 
{Adaptive optics (AO) systems represent the only possibility for studying the very dense environment of GC cores from the ground. 
We carried out observations of the core of \object{Terzan 5} in the near-IR bands with the ESO-VLT NAOS-CONICA instrument.} 
{We present the discovery of the likely counterpart in the \ks\ band and discuss its properties both in outburst and in quiescence. Archival HST
observations are used to extend our discussion to the optical bands.}
{The source is located at the blue edge of the turn-off area in the color-magnitude diagram of the cluster. Its luminosity increasefrom quiescence 
to outburst, by a factor 2.5, allows us to discuss the nature of the donor star in the context of the double stellar generation population of \object{Terzan 5} by 
using recent stellar evolution models.}

\date{Received 27 June 2012 / Accepted 25 September 2012}
\keywords{pulsars: general --
          pulsars: individual: IGR J17480-2446 --
          binaries: close --
          globular clusters: individual: Terzan 5}

\maketitle

\section{Introduction}
\object{IGRJ17480-2446} (hereafter \igr) was detected on October 10, 2010 with \textit{INTEGRAL} \citep{bordas2010} and 
then later identified as a newly discovered accreting pulsar with a spin period of 90.6 ms 
\citep{alta2010,stroh2010a}. 
The properties and evolution of its X-ray emission have been studied in detail \citep[see, e.g.][]{motta2011}, and the timing
analysis of the pulse period has revealed that the system has an orbital period of $\mathrm{\sim 21.3\ hr}$ \citep{stroh2010b,papitto2011}.
This is the most central pulsar in the globular cluster (GC) \object{Terzan 5}, located at about $\mathrm{4^{\prime\prime}}$ of the optical center.
The X--ray position has been accurately determined by \cite{pooley2010} with the satellite Chandra and, more recently, by \cite{riggio2012} using a
moon occultation method (MO) that allowed an accuracy of 0.04$^{\mathrm{\prime\prime}}$.

Low-mass X-ray binaries (LMXBs) are binary systems in which a compact primary (neutron star or black hole) accretes matter via mass transfer by Roche lobe 
overflow from a donor low-mass star. The accreting gas transfers angular momentum to the old neutron star, ``recycling" it to millisecond rotation periods
\citep{bhatta1991}.
This system is particularly interesting, because the  long spin period of the neutron star ($\mathrm{\sim 90.54\ ms}$) points to the possibility that it 
has been caught at the beginning of the mass-transfer phase that is accelerating it to a millisecond pulsar, thus it is a good candidate for enlightening  
the link between the X-ray sources and the radio millisecond pulsars proposed by this ``recycling scenario". \cite{papitto2012} analyze the 
spectral and pulse properties of \igr\ and find that the pulsar is indeed spinning up, and a similar result is also presented by \cite{patruno2012b}.

The detection and study of optical/IR counterparts of LMXBs is important in order to study
the nature and characteristics of donor stars and clarify the evolutionary status of the system. Detection of optical/IR counterparts in GCs is 
often difficult because of crowding conditions that affect many GCs. The discovery is usually achievable only from space or with adaptive 
optics (AO) systems from the ground. Near-IR bands have the advantage of a reduced extinction with respect to 
the optical bands, a crucial bonus for cases like this, in which the target cluster lies in a strongly obscured 
region \citep[$E(B-V) = 2.38$, meaning an extinction in the $V$ band of more than six magnitudes,][]{barbuy1998,valenti2007}. 

Terzan~5 is one of the densest and metal-richest clusters in our Galaxy \citep{cohn2002,ortolani2007}, and this characteristic
feature favored the formation of a large number of rotation-powered millisecond pulsars, with 35 pulsars discovered so far\footnote{see the web page maintained 
by P. Freire for an updated list: http://www.naic.edu/$\sim$pfreire/GCpsr.html} \citep{ransom2005,hessels2006,pooley2010}.
In addition, recent studies have revealed in Terzan~5  the presence of two different stellar populations: a metal-poor ($[Fe/H] \sim -0.2 \pm 0.1$) and a 
metal-rich one ($[Fe/H]\sim +0.3 \pm 0.1$)  \citep[][F09]{F09}. Although the presence of multiple populations is now recognized to be very common in 
GCs \citep[e.g.][]{carretta2009,piotto2009}, it is unusual to have such bimodality in metallicity. The two interpretations proposed \citep[see, e.g.][]{dantona2010}
are either that the two generations differ by $\sim$6~Gyr in age (F09), the younger being the most metal-rich, or they are almost coeval, but in this case 
the metal richer population must also show higher He content.
In this work, we report on the near-IR counterpart identification of \igr, obtained with observations taken after the detection of 
the outbursting pulsar and by comparison with archival data of the same field taken in the same bands and with similar techniques in 2008 by F09. 
The comparison allows us to determine the brightening of the source between the quiescent phase and the X--ray outburst of 2010, as well as 
its consequences for the binary model. In addition, the location of the source in quiescence with respect to the cluster average stellar loci
in the optical color-magnitude diagram allows us to consider its probable evolutionary status and to place it in the context of the two 
stellar generations that populate Terzan~5.
In Sect.~\ref{Photometry} we describe the data set and the procedure used for the detection of the IR counterpart, 
then we analyze the results with the help of photometry of archival optical data from HST and compare them with model predictions 
in Sect.~\ref{Analysis}. Finally, we briefly summarize the conclusions in Sect.~\ref{Conclusions}.

\section{Near-IR photometry\label{Photometry}}

\subsection{Observations, reductions, and calibrations}
Observations were carried during the X-ray outburst phase, within
a director's discretionary Time (DDT) proposal at ESO-VLT with the NAOS-CONICA AO
near-IR camera. Data were obtained during four nights in November 2010 in the $J$, $H$, and \ks\ bands,  
at the very beginning of the night at a considerably high airmass ($\mathrm{z \sim 1.9}$), owing to the limited 
visibility window of the target at the epoch of observation. Although in the $J$ and $H$ bands data were never good enough
to allow precise photometric analysis, the \ks-band images taken in nights 3 and 4
were of sufficient quality to resolve the very crowded center of \object{Terzan 5}.

The whole data set consists of 12 images of $\mathrm{2 \times 28s}$ each, which were reduced following
a standard procedure for infrared images: i) subtraction of a median sky frame, taken off-source
because of the crowding; ii) flat fielding; iii) alignment and
sum of all single images to obtain a final image of 672s equivalent exposure time.
On this image we used the DAOPHOTII \citep{stetson1987,stetson1994} version available in the suite IRAF\footnote{IRAF is distributed by 
the National Optical Astronomy Observatories, which are operated by the Association of Universities for Research
in Astronomy, Inc., under cooperative agreement with the National Science Foundation.} 
\citep{tody1986,tody1993} to detect and measure sources via both PSF profile fitting and aperture photometry. 
We used a semi-analytic model for the PSF, with a core described by an analytic function (in this case the best
choice among the available forms was a ``penny2'' function, i.e. a Gaussian core with Lorentzian wings model with
a beta parameter of 1.0, where the Gaussian core and Lorentzian wings may be tilted in different directions).
The outskirts were modeled by adding to the analytical function describing the wings of the PSF a nonanalytical 
contribution in the form of a residual correction matrix, established by using
all the stars used for the PSF model, and weighing the contributions by the inverse square of the noise. 
In this way we found that the PSF model varies with the distance from the star chosen for the wavefront analysis with a 
trend described well by a quadratic function of the radius. 
Calibration was obtained via comparison with the photometry of F09 obtained on the same field
with the Multi-conjugated Adaptive Optics (MAD) camera at ESO-VLT, by analyzing a subset of stars in common 
with our observation after selecting those that were not too bright to prevent systematic effects of nonlinearity.

\begin{figure*}
 \centering
\includegraphics[width=17cm]{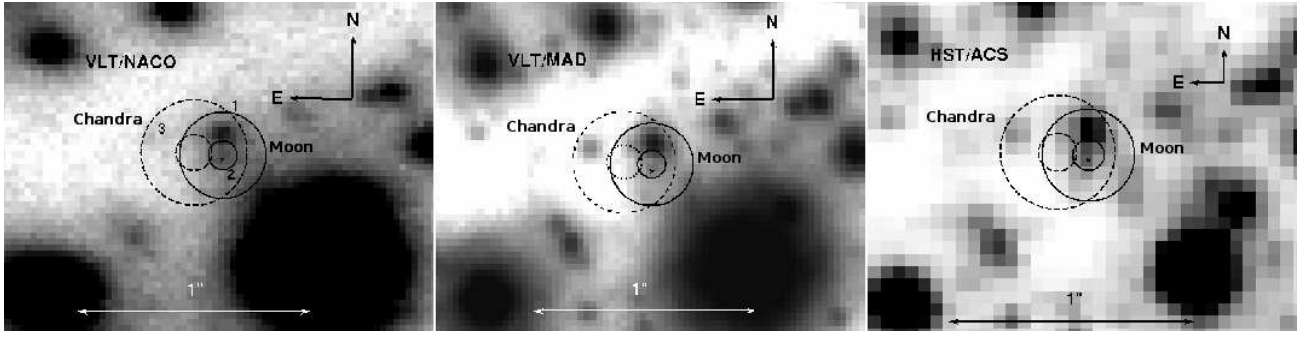}
 \caption{Map of 1.5$^{\prime\prime} \times$1.5$^{\prime\prime}$ area surrounding the position of \igr. 
Left panel: VLT/NACO image (this work), with superimposed error circles
drawn at 1$\mathrm{\sigma}$ and 3$\mathrm{\sigma}$ radii of the X-ray transient positions obtained from Chandra (dashed circles) 
and the Moon Occultation method (solid circles). 
X-Ray transients positions are computed combining the errors given by the authors (0.04$^{\prime\prime}$ for MO, 0.06$^{\prime\prime}$ for Chandra) 
and the uncertainty on the astrometric fit (0.06$^{\prime\prime}$).
Central panel: VLT/MAD image. Right panel: HST/ACS image in the F606W filter. The three sources inside the error boxes
are marked as well. Source \#2, marked with a dot, is the candidate counterpart (see text).}
 \label{fig:map}
\end{figure*}

\begin{table*}
 \begin{tabular}{c c c c c c c }
\hline
 ID\# & R.A. & Dec. & K(2010) & $\sigma(K)_{2010}$ & K(2008) & $\sigma(K)_{2008}$ \\
\hline
  1   & 17:48:04.822 & -24:46:48.81 & 13.88 & 0.05 & 13.91 & 0.02 \\
  2   & 17:48:04.822 & -24:46:48.90 & 15.40 & 0.15 & 16.44 & 0.15 \\
  3   & 17:48:04.845 & -24:46:48.82 & 16.10 & 0.40 & 15.90 & 0.10  \\
\hline
 \end{tabular} 
\caption{Photometry of the sources found within 3$\sigma$ of the X-ray transient
position (both MO and Chandra).}
\label{tab:sources}
\end{table*} 

\subsection{Astrometry, photometry, and counterpart identification}

To obtain the astrometric calibration, we attempted to use the UCAC3 catalog \citep{zacharias2010}, which is an improved
version of the more commonly used USNO catalog obtained with astrographic techniques, but we found
only one catalog star in the small field of view of NACO. 
We therefore followed a two-step approach, using archival images from the HST archive. 
In particular, we used two images of \object{Terzan 5} taken with the ACS/WFC in 2003 
in the F606W (R) and F814W (I) filters. These images already have an astrometric solution 
written in the image header, which was found to have a 0.5$^{\prime\prime}$ offset from the UCAC3 reference positions.

As a first step, we thus obtained an astrometric solution by using 21 UCAC3 catalog stars present in the 
field of view of HST/ACS, with a magnitude range between 11 and 16. 
Then, we used the sources from HST images as secondary astrometric calibrators for our VLT/NACO images.
As a byproduct, we also obtained the photometric catalog of the stars in the HST/ACS images.
The uncertainty on the astrometric fit, obtained as described above, was estimated by combining the error from the
first fit and the fit of the secondary calibration process. For the first we assumed as source errors the r.m.s. of the fit, 
the positional error on the catalog given in the UCAC3 table, the centroiding error on the HST stars, and the systematic
error of the UCAC3 catalog with reference to the absolute coordinate system. 
The secondary astrometric fit has a negligible error (formal r.m.s. of 0.01 m.a.s.) so that we neglect it in the total
error estimation. The adopted uncertainty on the astrometry for the our VLT/NACO position is 50 m.a.s., on the basis
of the analysis described above.

Figure \ref{fig:map} (left panel) shows a $\mathrm{~1.5^{\prime\prime} \times ~1.5^{\prime\prime}}$ subimage around 
the X-ray transient position. The positions of the X-ray transient  found with
the Moon occultation method and with Chandra are shown with error circles
drawn at 1$\mathrm{\sigma}$ and 3$\mathrm{\sigma}$ radii, computed by combining the errors quoted by the authors 
(0.04$^{\prime\prime}$ for MO, 0.06$^{\prime\prime}$ for Chandra) and the uncertainty on the astrometric fit 
(0.05$^{\prime\prime}$). 

We found three sources within 3$\sigma$ radius of the X-ray transient positions.
For the MO position, the easternmost source (\#3 in Fig. \ref{fig:map}) is outside the error box, so we only consider
it as a possible counterpart to the Chandra position. 
Table \ref{tab:sources} reports the position and magnitudes of the three sources.
Source \#2, on the other hand, is very close to the nominal position of the MO X-ray transient. If we consider the latest
measurement performed by using the lunar laser altimetry by the LRO/NASA satellite \citep[see][]{riggio2012}, the two 
positions differ by 32 m.a.s..
To determine whether one of the three candidate sources had varied with respect to the epoch of MAD observations, 
we compared our catalog with the one from F09. One of the sources (\#2) is not 
present in the F09 catalog. This source has magnitude $K_{\mathrm{S}}=15.40 \pm 0.15$ (see Table \ref{tab:sources}),
whereas the limiting magnitude of the VLT/NACO image is $K_{\mathrm{S}}=16.2$ (at $3\sigma$ level).

We checked in the images of F09 that the source is indeed visible but
not included in the photometric catalog, or if it is not visible at all. 
We therefore selected the best K-band images from the MAD dataset (i.e. 
those having the lowest values of FWHM) and again performed the photometry in the subimage around the 
X-ray transient position. In this image, the object has been detected and measured by DAOPHOTII,
finding $K_{\mathrm{S}} = 16.44 \pm 0.15$, with a $3\sigma$ level magnitude limit of $K_{\mathrm{S}}\sim19$,
while the other two sources do not show variation in magnitude between the two epochs.

On the basis of the results described above, it can be seen that the source has increased its brightness by $\sim$1.04 mag 
between the two epochs, i.e. a factor of $\sim 2.5$, and thus we identify it as the most probable IR counterpart to the X-ray transient \igr. 
Figure \ref{fig:map} (central panel) shows the same area around the transient position on the MAD image, 
while the corresponding area in the HST/ACS F814W image is shown in the right hand panel of the same figure.
According to our astrometric calibration, the position of the source is
$\mathrm{\alpha}$=17:48:04.822 ; $\mathrm{\delta}$=-24:46:48.90, to which we associate 
an 1$\sigma$ uncertainty of 0.05$^{''}$. 

\begin{figure}
 \centering
\includegraphics[width=10cm]{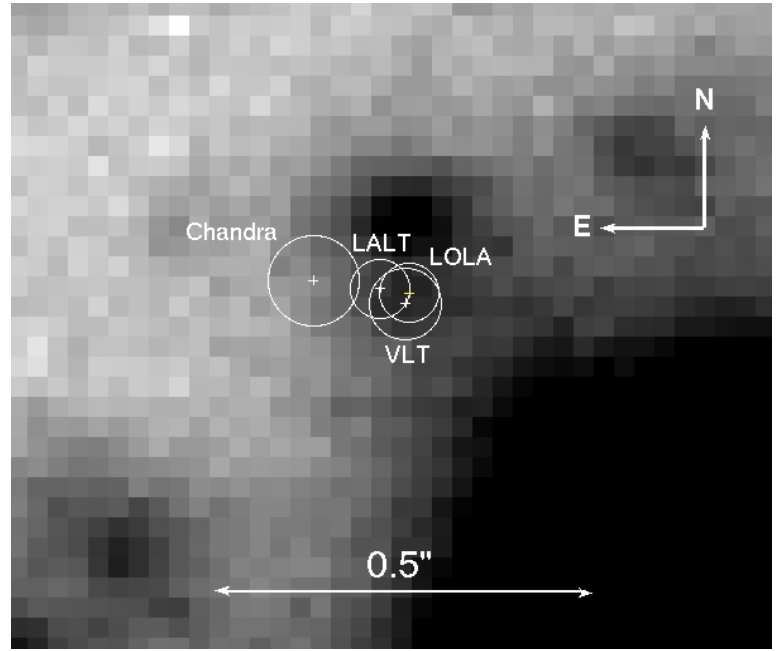}
 \caption{Map of the area around the pulsar \object{IGRJ17480-2446}. Circles with a cross at the center mark the four 
transient positions as determined by \cite{pooley2010} (Chandra) and \cite{riggio2012} with the Moon occultation 
method (LALT and LOLA) and this work (VLT).}
 \label{fig:positions}
\end{figure}

Figure \ref{fig:positions} shows the X-ray transients positions determined with Chandra \citep{pooley2010} and with
Rossi/XTE and the Moon Occultation method, using two different laser altimetry onboard lunar missions, i.e. LRO/NASA (LOLA)
and NAGUYA/JAXA (LALT) \citep{riggio2012}, and the IR counterpart position discussed in this work.

\subsection{Comparison with archival HST observations during quiescence}
Another check was made by using the two archival images taken with the HST/ACS in the F606W and F814W filters.
The source is visible in the two HST images, although the image is undersampled and the crowding conditions extreme.
Despite the undersampling (FWHM $\sim$ 1.8 pixels), DAOPHOTII was able to detect
and measure the candidate, although the procedure was more challenging and required a manual identification of the
source. We found $F606W = 23.52 \pm 0.23$, $F814W = 20.65 \pm 0.19$ for the source in the 
Vegamag system. 

\begin{figure}[t]
\includegraphics[width=8cm]{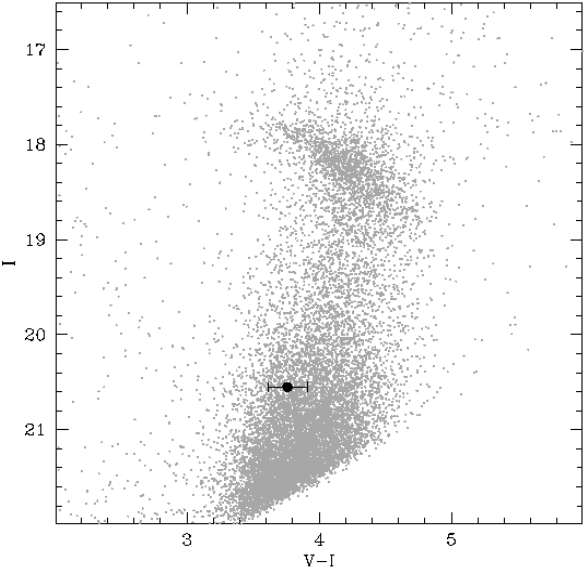}
 \caption{$I,V-I$ color-magnitude diagram of the central area of \object{Terzan 5} from HST/ACS data. The candidate counterpart is marked with a
large black dot.}
 \label{fig:hstcmd}
\end{figure}

We then transformed the magnitudes of the HST/ACS catalog into standard V,I magnitudes by using 
the \cite{sirianni2005} relations, and obtain $V = 24.31 \pm 0.22$, $I = 20.55 \pm 0.20$
for the candidate counterpart,.
Figure \ref{fig:hstcmd} shows the color-magnitude diagram with the
position of the candidate counterpart superimposed. The mean loci of the color-magnitude diagram
are considerably spread out most probably because of the differential reddening affecting the cluster.

We compared our results with \citet{patruno2012}, who used the same 
HST archival data set as used in this work. The authors report identifying the optical companion of our IR counterpart
as their star \#2, which has magnitudes and colors that are consistent with the ones reported above. However, star \#2 of \cite{patruno2012}
does not seem to correspond to our proposed conterpart. It is a faint object located halfway between the 
bright star just above the counterpart \citep[\#4 in ][]{patruno2012} and the much brighter star located to the SW. 
It is at about 0.6$^{\arcsec}$ from the X-ray source position as determined by Chandra and the Moon occultation observed by RXTE.
This object is not visible in the IR image of Fig. \ref{fig:map} because the choice of the contrast level is such that it is embedded in the PSF 
of the bright star, but it is visible in the HST/ACS image (see right panel of Fig. \ref{fig:map}).
However, it has been detected and measured in the near-IR, showing no sign of variation between the 2008 and 2010 epochs. Moreover, 
in the optical images it looks extended, and in the F814W image it has been actually deblended into 
two separate components. On the basis of the discussion above, we conclude that the counterpart proposed in \cite{patruno2012} does not correspond 
to our IR counterpart, but it is another object that has similar magnitude and color.

\section{Analysis and comparison with theory\label{Analysis}}

The counterpart, at the epoch of HST observations (2003) is presumably in quiescence
and lies in the turn-off region of the color-magnitude diagram, slightly bluer than the average main sequence
locus. We take it as a bona--fide position, although it could be simply due to the relatively high 
photometric error and/or to the differential reddening affecting the cluster. 
Considering the differential reddening, we note that the average value for reddening in \object{Terzan 5} is $E(B-V) = 2.38$
\citep{barbuy1998}, and recent determinations indicate that 
in the source area, very close to the optical center of the cluster \citep{massari2012}.
This correction does not help put the object closer to the main sequence mean locus, because the source would be
even bluer in a dereddened diagram. We therefore assume that the star is actually on the blue edge of the main sequence 
locus at the turnoff level.

We now make a differential analysis of the few data we have on this system, to understand whether they 
form a coherent frame in which to interpret its evolution and properties. We try to answer to two main questions:
\begin{enumerate}
 \item Are the location of the optical companion in the HR diagram in 2003 and the increase by a factor $\sim$2.5 between 2008 and 2010 in its IR luminosity
consistent with the system's evolutionary status?
  \item Can the photometric results for the optical counterpart provide indications that attribute \igr\ to one of the two stellar populations of Terzan~5?
\end{enumerate}    

\subsection{X--ray heating during the X--ray transient phase}
First of all, the location of the object identified as a possible companion of the pulsar, close to the cluster turnoff 
during the quiescence status of 2003, shows that its mass must be close to the turnoff mass in this cluster. 
We put it here at \msec=0.8\msun, but see the discussion in Sect.~\ref{Elio}. We choose a typical neutron star mass of 
\mprim=1.4\msun\ for the pulsar (see the justification below). The X-ray burst activity of the source and the fact that the 
neutron star has been accelerated up to the 90 ms spin period indicate that the neutron star is accreting from the donor, 
so that the donor radius must be close to its Roche lobe radius. We can therefore require that the companion
radius R$_2$\ is equal to its Roche lobe radius $R_{\mathrm{RL_2}}$, where $R_{\mathrm{RL_2}}$ can be calculated using the standard
approximation by \cite{paczynski1971}.

From similar optical studies performed on other known accreting LMXBs we know that the observed optical counterpart of 
these sources is too bright to be the result of intrinsic emission from the companion star (expected to be a low-mass star,
usually less than 1 \msun). Optical emission during X-ray outbursts comes from the X-ray irradiated face of the companion 
star and/or the accretion disk \citep[see e.g.][]{giles1999,wang2001}. Indeed, the 
optical luminosity is observed to decay with time as the X-ray flux decays during the outburst, until it reaches a minimum 
value during X-ray quiescence. Both the disk and one face of the companion will be irradiated by the X-ray emission from 
the central source.

The X-ray luminosity - during the outburst in the date of IR observations - was $4.3 \times 10^{37}$ erg/s \citep{papitto2012}, 
calculated from the measured X-ray flux and spectrum of the source, at a distance of 5.9 kpc, which is the 
current estimate for \object{Terzan 5} \citep{lanzoni2010}. In the hypothesis that this luminosity heats the accretion disk 
and a side of the companion star and assuming isotropic emission, we can evaluate the fraction of the irradiation luminosity 
that will be intercepted and reprocessed by the disk and the companion star, respectively \citep[as, e.g., in][]{burderi2003,campana2004}. 
For the accretion disk, the fraction of intercepted luminosity, $f_\mathrm{D}$, is given by the projected area of the disk as seen by the 
central source, $2 \pi R \times 2 H(R)$ (where $R$ is the disk outer radius and $H(R)$ is the disk semi-thickness at $R$) divided by the 
total area, $4 \pi R^2$. Adopting a standard Shakura-Sunyaev \citep{shakura1973} disk model, 
in which the outer edge is truncated at 80\% of the Roche lobe of the neutron star, the orbital period of $\sim 21$ hr, a 
Roche lobe-filling companion star of about 0.8 \msun, we find $f_\mathrm{D} \simeq 4.7 \times 10^{-2}$. For the companion star the 
intercepted fraction can be written as $f_\mathrm{C} = 2 \pi a^2 (1 - \cos \theta) / (4 \pi a^2)$, where $a$ is the orbital separation and 
$\theta$ the angle subtended by the companion star as seen from the central source. In the same hypothesis as 
above, we obtain $f_\mathrm{C} \simeq 2.5 \times 10^{-2}$.

Adopting an X-ray albedo of 0.85 and 0.93 for the companion star and the accretion disk, respectively \citep[corresponding to 
an averaged X-ray albedo of $\sim 0.9$, see e.g.][]{vanparadijs1996}, this corresponds to similar 
luminosities of the irradiated face of the companion star and the irradiated disk, $L_\mathrm{C} \sim L_\mathrm{D} \sim 1.4 \times 10^{35}$ 
ergs/s, respectively. 
The corresponding reprocessed blackbody temperatures are estimated to be $T_\mathrm{C} \sim 11400$ K for the companion star 
and $T_\mathrm{D} \sim 11980$ K for the irradiated surface of the disk. From the two blackbodies of temperatures $T_\mathrm{C}$ and $T_\mathrm{D}$
and luminosities $L_\mathrm{C}$ and $L_\mathrm{D}$, respectively, we have calculated the predicted apparent \citep[incremented for the 
appropriate extinction correction towards \object{Terzan 5},][]{cardelli1989,valenti2007} magnitudes in three IR bands,
 $J$, $H$, and $K_\mathrm{S}$, which are $m_J \simeq 17.5$, $m_H \simeq 16.7$, and $m_{K_\mathrm{S}} \simeq 16.2$ 
for the irradiated face of the companion star and $m_J \simeq 17.3$, $m_H \simeq 16.6$, and $m_{K_\mathrm{S}} \simeq 16.0$ 
for the irradiated disk, respectively. Combining the $K_\mathrm{S}$ magnitudes of the irradiated companion star and accretion
disk, we estimate a magnitude of $m_{K_\mathrm{S}} \sim 15.3$ for the observed IR counterpart, which is compatible with the observed 
$K_\mathrm{S}$ magnitude of our candidate. The largest uncertainty in our calculation of the irradiated luminosity is in the
X-ray albedo. In principle, if in the future it were possible to disentangle the companion star and accretion disk 
reprocessed luminosities, it would be possible to put a constraint on the X-ray albedo of these two components of the
system.

\begin{figure}[t]
\begin{center}
\includegraphics[width=8cm]{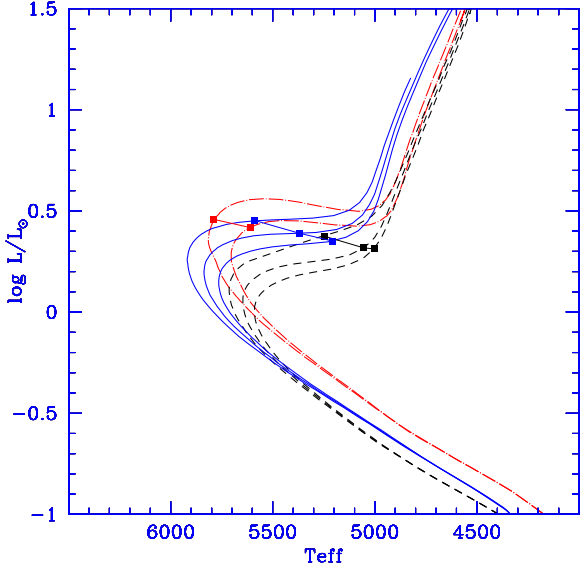}
\caption{Isochrones representing the two possible stellar populations in \object{Terzan 5}. The basic population has
Y=0.26 and Z=0.01 (blue full lines, isochrones of 10, 12, and 14Gyr). The second population may either be coeval (dashed black lines, 
same isochrones for Y=0.33 and Z=0.03) or be much younger (red dash--dotted isochrones for 6 and 8 Gyr,  Y=0.29, Z=0.03). 
The lines with squares represent the location for each set of isochrones at which the evolving mass has a radius equal to the Roche lobe 
radius of \igr\ in a system with orbital period P=21.274hr.
}
\label{f3}
\end{center}
\end{figure}  

\subsection{Which population for \igr? \label{Elio}}

In the 2003 observations, the optical companion must instead be close to its intrinsic luminosity. In principle, when accretion onto the neutron 
star is off, the neutron star should still emit energy according to the Larmor formula because it is a rotating magnetic dipole, and part of 
this energy can be intercepted and reprocessed by the companion star. Indeed, this mechanism has been invoked to explain the high optical luminosity of the
companion star of the accreting millisecond pulsar SAX J1808.4-3658 in quiescence \citep{burderi2003,campana2004}, and see \citet{davanzo2009}
for similar studies on other sources of this class.

To estimate this effect in the case of IGR J17480-2446, we should start from the estimates of the neutron star magnetic field. \citet{papitto2011}
evaluate the magnetic field in the range between $\mathrm{ \sim 2 \times 10^8} $ G to  $\mathrm{\sim 2.4 \times 10^{10}} $ G, assuming that the inner 
disk radius lies between the neutron star radius and the corotation radius, while the source shows pulsations. 
\citet{papitto2012} estimate the magnetic field in the range between $\mathrm{ \sim 5 \times 10^9 }$ G and $\mathrm{ \sim 1.5 \times 10^{10}} $ G 
from the derived estimate of the NS spin up, which is $\mathrm{\dot \nu \simeq 2.7 \times 10^{-12}}$  Hz s$^{-1}$, attained when the X-ray 
luminosity reached its peak value and assuming that the NS accretes the Keplerian angular momentum of matter at the inner disk boundary. 
These estimates agree with those given by \citet{cavecchi2011}, who estimate
the magnetic field in the range $\mathrm{2 \times 10^8}$ G to $\mathrm{3 \times 10^{10} }$ G, using arguments similar to \citet{papitto2011}.
Finally, \citet{miller2011} use the results of a relativistic iron line fit to estimate the magnetic field at the poles to be 
$B = \mathrm{9(2) \times 10^8} $ G or $B =\mathrm{  3(1) \times 10^9} $ G depending on the assumed value (1 or 0.5, respectively)
of the conversion factor from spherical to disk accretion when balancing magnetic and ram pressures. In agreement with all these estimates,
we consider an upper limit to the neutron star magnetic moment in this system of $\mathrm{ \mu < 2.4 \times 10^{28}}$ G cm$^3$ and the spin frequency 
of the neutron star $\mathrm{\nu \simeq 11.045}$ Hz.

With these values, the Larmor formula gives an emitted power of $L_{\mathrm{PSR}} \le 3.3 \times 10^{32}$ erg/s, resulting in a heating luminosity
of $L_{\mathrm{h}} = f\  3.3 \times 10^{30}$ erg/s, which is about $2.5 \times 10^{-3}$ times lower than the intrinsic bolometric luminosity of a
0.8 \msun\ companion. We therefore conclude that, in the case of \igr, the optical luminosity in quiescence is truly 
representative of the intrinsic luminosity of the companion star.

We can now try to interpret the location of the companion close to, but bluer than, the turnoff of the cluster. 
We first recognize that, independent of how long ago the system was formed, the pulsar acceleration began a short time ago, and it is to be attributed to the
 evolution of the donor to larger radii. This recalls the evolutionary status of the system PSR J1740-5340 in NGC 6397, whose very fast radio millisecond 
pulsar was indeed accelerated when the companion began its  evolution off-main sequence, as shown by the models presented in \cite{burderi2002}.
In PSR J1740-5340  today the companion mass has been reduced to $\sim$0.4\msun, whereas in \igr\ we must be close to the beginning of a similar evolutionary path. 
The evolution -- when irradiation is low -- proceeds from the turnoff towards the red giant branch, so that we should expect that the donor star 
is in fact {\it redder} than the turnoff. Its bluer location might mean that it suffers a low illumination from the pulsar
 \citep[see Fig.~1 in][left side evolution]{burderi2002}, but, as mentioned above, we find this explanation unlikely.

We now examine this problem from a different point of view, by using a detailed model for the Terzan~5 cluster. 
Recent observations of the color-magnitude diagram features and the chemistry of \object{Terzan 5} constitute a benchmark in our understanding of GC formation. 
F09 show that the cluster clump (horizontal branch) stars are actually divided into two clumps separated by $\delta M_\mathrm{K} \sim 0.3$ mag and 
that the more luminous stars have a much higher iron content ($[Fe/H] \sim +0.3 \pm 0.1$) with respect to the lower HB ($[Fe/H] \sim -0.2 \pm 0.1$). 
This result shows that two well-separated populations are present in the cluster.  Comparing the HB data to stellar isochrones of the correct metallicity, 
F09 conclude that the population with a higher metallicity must be $\sim$6~Gyr younger than the first generation. The presence of such a young
population among bulge stars and clusters is not seen elsewhere \citep[e.g.][]{feltzing2000, origlia2008}, and \cite{dantona2010} propose instead 
to attribute the different clump luminosity to a much greater helium content present in the metal-rich population, consistent with models for the chemical 
evolution of the bulge \citep[e.g.][]{renzini1999}.

In Figure~\ref{f3} we show isochrones from \cite{dantona2010}, able to describe the two populations in Terzan~5 and the two different cases 
(age {\it vs} He-content difference between the two populations). 
We report, on the isochrones, the location of the evolving star when the evolving mass has a radius equal to the Roche lobe radius of \igr\ in a system 
with orbital period P=21.274 hr. 
Despite the (small) mass loss that the donor must have suffered already, the theoretical location should still be close to the location of the donor of \igr. 
Since the observations show that the donor in quiescence lies at the blue edge of the turn-off locus, the comparison between the possible locations 
implies that either the donor belongs to the low-metallicity population (solid lines, blue in the color electronic version) and the 
second population is helium rich (dashed lines, black in the color version), or that it belongs to a younger second population (dot--dashed lines, red in the color 
version). Only in these two cases may we expect to find it at colors bluer than the turnoff location.

\section{Summary and conclusions\label{Conclusions}}

In this work we present and discuss the likely detection and characterization of the near-IR counterpart to
the X-ray pulsar \object{IGRJ17480-2446}. The analysis shows that the source is visible in quiescence both in near-IR 
and optical bands. Thanks to the precise identification of the optical counterpart, we were able to 
trace its location back in the color magnitude diagram to the previous epoch of quiescence. 
The relative location of the object with respect to the ridge line of the cluster turnoff offers us the intriguing 
possibility of attributing the object either to the first (low-metallicity) population, if the second population is 
also helium rich,  or to the second one (high-metallicity), if it is much younger than the first one (Fig.~\ref{f3}).
A fundamental test to determine which population \igr\  belongs to would be to determine the metallicity,
which seems currently out of reach with the present generation of telescopes and instruments available.
Determining metallicity will allow to choose between the two scenarios (age or He-content difference) proposed to explain the different
luminosities of the clump stars of these two populations. 


\begin{acknowledgements}
We gratefully acknowledge the excellent support and help from the ESO staff. 
AP acknowledges the support by the grants AYA2009-07391 and SGR2009-811, 
as well as the Formosa program TW2010005 and iLINK program 2011-0303.
\end{acknowledgements}

\bibliographystyle{aa}
\bibliography{paper_new}


\end{document}